\documentstyle[11pt,aaspp4]{article}

\slugcomment{To Appear in November 1997 PASP}

\lefthead{Gizis \& Reid}
\righthead{M Subdwarf Secondaries}

\begin{document}

\title{M Subdwarf Secondaries: A Test of the Metallicity Scale}
\author{\sc John E. Gizis \& I. Neill Reid}
\affil{Palomar Observatory, 105-24, California Institute of Technology,
  Pasadena, California 91125, e-mail: jeg@astro.caltech.edu, 
  inr@astro.caltech,edu}

\begin{abstract}
We present spectra of three M subdwarfs which are common proper motion 
companions 
to F or G subdwarfs of known metallicity.  The assumption that the 
companions have the same composition allows us to test the
Gizis (1997, \aj, 113, 806) 
M subdwarf classification system and its correspondence
to metallicity.  
The results are in excellent agreement
with the Gizis (1997) scale, thereby 
showing that the Allard \& Hauschildt
(1995, \apj, 445, 433) Extended model atmospheres agree well
in the 6200 -- 7400$\AA$ region for cool metal-poor stars.  
We also show that the results are consistent with the 
main sequences of globular clusters using the 
Reid (1997, \aj, 114, 161) distance scale.
\end{abstract}


\section{Introduction\label{intro}}

The metal-poor stars of the thick disk and halo provide 
an invaluable record of Galactic history.  The main-sequence
FGK subdwarfs have proven to be an important source of information
on these populations (e.g., \cite{carney}). 
The much cooler and fainter M subdwarfs offer an important
alternative tracer group.  Indeed, in addition to the possibility of
observing nearby proper motion M subdwarfs, it is now feasible to 
obtain both photometry for the M subdwarfs in globular clusters
with the Hubble Space Telescope (e.g., \cite{seg96}) 
and spectra for M dwarfs 
and M subdwarfs at distances of a few kiloparsecs above the galactic plane
with 10-meter class telescopes (\cite{keck}).  
Using these objects as probes of Galactic structure, however, 
requires a good understanding of their 
properties in order to derive metallicities and luminosities. 

Gizis (1997, hereafter G97) has presented a spectroscopic classification scheme
which is based on  moderate resolution ($\sim 3 \AA$) spectra covering the
wavelength range 6200 -- 7400 $\AA$.  Quantitative bandstrength
indices measuring TiO and CaH features are used to classify
stars as M V (ordinary disk stars), sdM (M subdwarfs), and esdM
(extreme M subdwarfs).  Comparison to the Allard \& Hauschildt
(1995) synthetic spectra allowed G97 to show that these classes
correspond to $[m/H] \sim 0.0$, $[m/H] \sim -1.2 \pm 0.3$,
and $[m/H] \sim -2.0 \pm 0.5$  respectively.  Comparison of
the ($M_V,V-I$) HR diagram shows that HST globular cluster
sequences (\cite{seg96}) and stellar interior calculations 
with Allard \& Hauschildt (1995)  model atmospheres (\cite{bcah95}) are in agreement with this
scale; however, serious systematic errors could in principle affect
all of these methods of estimating metallicity.  

We present spectra of three M subdwarfs which are companions to 
hotter subdwarfs of known metallicity.  Our aim is to test the 
metallicities derived from the M subdwarf spectra 
by comparison with those measured for their better understood
primaries.  The data are presented in 
Section~\ref{data}, the implications for the metallicity scale
are discussed in Section~\ref{discuss}, and the results are
summarized in Section~\ref{conclusions}.  

\section{Observations \label{data}}

We observed two systems (G116-009; G176-046) 
whose low-luminosity components were recently
discovered (\cite{mr92}; \cite{mrzo95}).  
We also observed VB 12 (LHS 541), the low luminosity companion
to HD 219617 (LHS 540) discovered by van Biesbroeck (1961).  
All three stars were identified as common proper motion companions
to already-known, relatively-bright high proper motion objects.  
We refer to the low-luminosity component of G 116-009 
as G 116-009B and the primary as G 116-009A. 
CLLA find that G 116-009A has $B-V = 0.86$, 
corresponding to $T_{eff} = 4750$ K.
The case of the G 176-046 system is rather complicated ---
Martin et al. (1995) note that their object
is the fourth member of this system, since 
Latham et al. (1992) deduce that G 176-046 is a spatially unresolved triple
from their high-resolution spectra.  
We therefore will refer to the low-luminosity companion as G 176-046D
and the primary as G 176-046ABC.  Latham et al. (1992) find
that this ``primary'' has $B-V = 0.80$ and $T_{eff} = 4860$ K.   
In fact, this system is a quintuple system, since
Ryan (1992) has shown that LP 215-35 is a common
proper motion companion at $343\arcsec$ separation.
This companion has $B-V = 0.86$.  
VB 12's primary, HD 219617, is itself a double star
with a orbital semi-major axis of $0.80\arcsec$ (\cite{h91}).  
CLLA find $B-V = 0.48$ and $T_{eff} = 5857$ K.  

The stars were observed with the Palomar 200-in on UT Date 
1 June 1997 using the Double Spectrograph.  A dichroic which
divided the light at $5500 \AA$ was used.  
The red camera was used with a 600 l/mm grating, yielding
wavelength coverage from 6040 to $7380 \AA$ at $\sim 3 \AA$ resolution.  
Very few counts were obtained in the blue camera
for the M subdwarfs and those data were therefore not used.
The setup is similar to that used by   
Reid, Hawley, \& Gizis (1995) and G97.   
The spectra were wavelength calibrated with neon and argon lamps
and flux calibrated with the Gunn \& Oke (1983) standards using
FIGARO.  

\begin{figure}
\figurenum{1}
\plotone{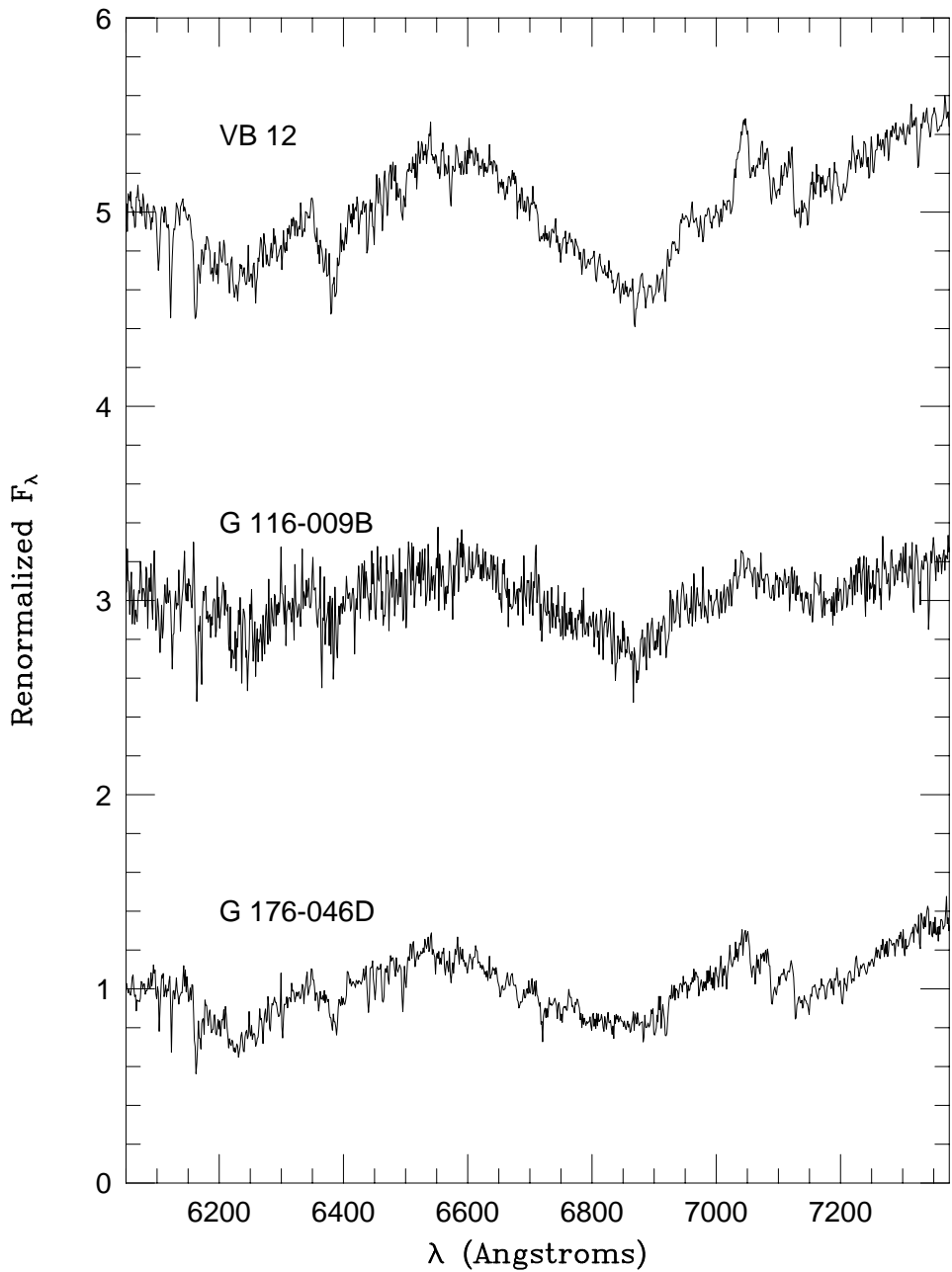}
\caption{Spectra of the three metal-poor M subdwarf companions.  
The spectra have been renormalized by a mean value of 1.
G116-009B and VB 12 respectively have been displaced upwards by 
2 and 4 respectively.  
\label{fig-spectra}}
\end{figure}

The resulting spectra are plotted in Figure~\ref{fig-spectra}.
We measure bandstrength indices defined in Table~\ref{indices} 
as the ratio of flux in the features (W) to flux in the 
pseudo-continuum regions (S1 and S2).  They were originally
defined in Reid et al. (1995).  Our measurements of the indices 
as well as photometry from the literature 
are reported in Table~\ref{table-data}.
Standards from Marcy \& Benitz (1989) were
used as templates to determine radial velocities
accurate to $\pm 20$ km s$^{-1}$ for the M subdwarfs.
All are consistent with the more precisely-known primary velocities.   
The metallicities derived by Carney et al. (1994, 
hereafter CLLA) for the primaries are also listed in Table~\ref{table-data}.  

\section{The Metallicity Scale\label{discuss}}

\begin{figure}
\figurenum{2}
\plotfiddle{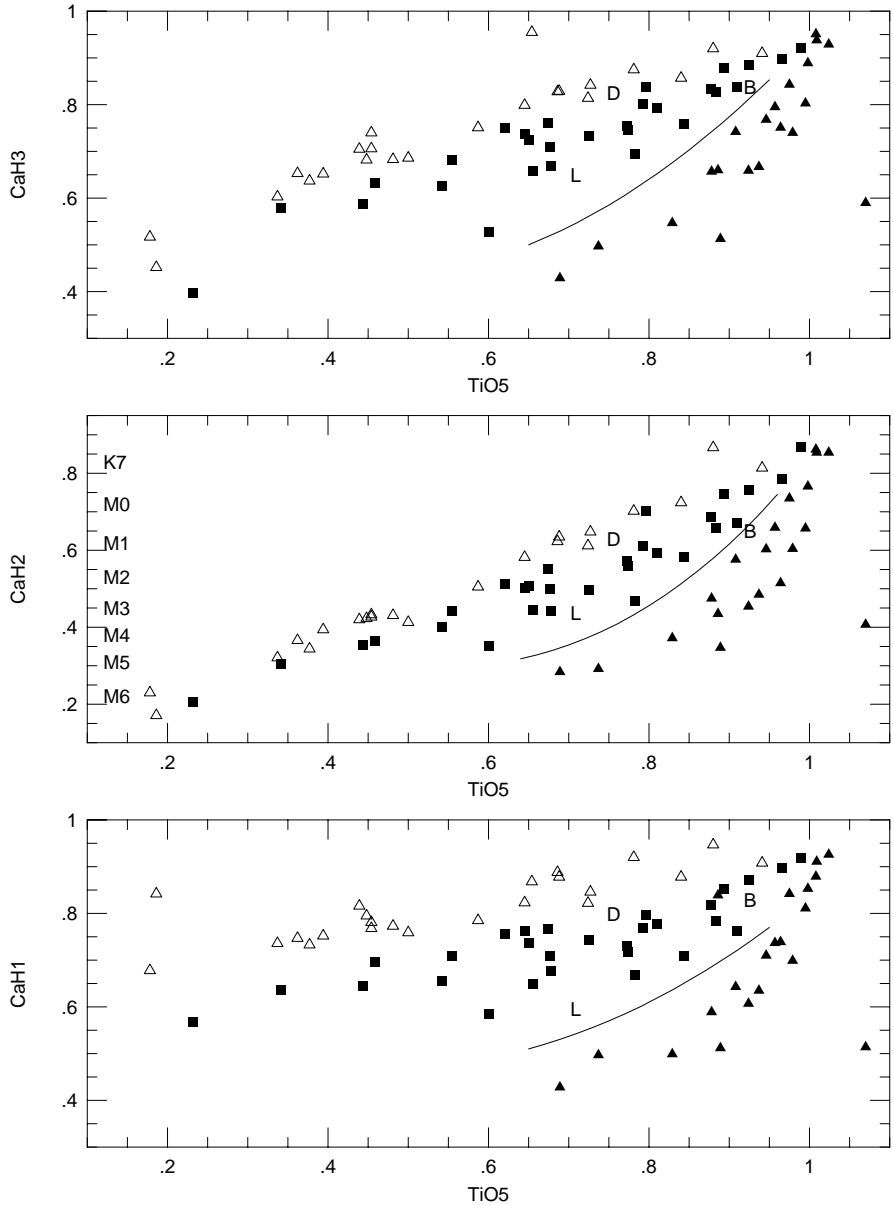}{6in}{0}{70}{70}{-230}{-30}
\caption{The TiO and CaH indices of the three low-luminosity
companions (solid circles) are plotted with the Gizis (1997) values for
high velocity M V (open triangles), 
sdM (solid squares), and esdM (solid triangles). 
The solid curves show the approximate boundary between 
sdM and esdM.   
LHS 541, G 116-009B , and G 176-046D are plotted as 
L, B, and D respectively.
G176-046D lies close to the M V/sdM boundary. 
The sdM G 116-009B and VB 12 lie just above the esdM 
sequences.  \label{fig-tiocah}}
\end{figure}

G97 found that the mean $[m/H]$ for the 
sdM and esdM are $-1.2$, with a range of $\pm 0.3$, and 
$-2.0$, with a range of $\pm 0.5$, respectively.
G97 also argued that stars of $[m/H] \gtrsim -0.6$ are
not distinguishable on the basis of their indices from ordinary 
(near solar metallicity) nearby M dwarfs.
The three stars in the present sample have classifications
of M1.0 V, sdM0.5, and sdM3.0.  We can derive more quantitative 
metallicity estimates by considering the values of
bandstrength indices rather than the shorthand classification.
The three stars are compared to the G97 standards 
in the TiO-CaH diagrams in Figure~\ref{fig-tiocah}.  The two sdM
stars G 116-009B and VB 12  
lie quite close to separation line between sdM and esdM, implying 
that they have similar $[m/H]$ near to the lower end of the sdM range, i.e.
$-1.4$ or $-1.5$.  The indices of G 176-046D indicate that it
is substantially more metal-rich than the other stars.
Although classified as ``M1.0 V'', G 176-046D   
lies within 0.01 in CaH1  of being classified as ``sdM.''  This offset 
is less than the observational error of $\pm 0.02$.  
We conclude that this subdwarf lies at the upper range of sdM abundances
and at the lower extreme of the high-velocity disk 
(Intermediate Population II) stars, so  $-0.6 < [m/H] \lesssim -0.9$.  

These expectations from the analysis of the M subdwarf spectra are confirmed 
by the CLLA measurements of the primaries.  
G 116-009A has $[m/H] = -1.46$\footnote{The metallicities 
for G 116-009A and G 176-46ABC cited in
Martin et al. (1995) are those derived by  Laird, Carney, \& Latham (1988) 
which are significantly more metal-poor than the more recent
estimates given by CLLA.  As a result, the comparison of G176-046 
and G116-009 to $\omega$ Cen made by Martin et al. 
is no longer appropriate.}.  
This system provides the cleanest test, since CLLA's 
finding of no radial velocity variations implies the primary
is single.  HD 219617 has $[m/H] = -1.40$ but is an unresolved binary.
It should be noted that Axer, Fuhrmann, \& Gehren (1994) derive a 
significantly higher value of $[Fe/H] = -1.08$, but specifically 
note that their value is ``suspect.''  The CLLA
values are in good agreement with our M subdwarf estimates
for both stars above.  
Finally, we must consider G 176-046.  CLLA 
derive $[m/H] = -1.07$ but since the ``primary'' is
made up of three stars this value is uncertain.  An independent
estimate may be obtained from the Ryan (1992) UBV photometry of the 
distant subdwarf companion LP 215-35.  Inspection of Ryan's
Figure 1 shows that LP 215-35 is more metal-rich 
than $[m/H] = -1$.  His photometry of G 176-046ABC 
implies $[m/H] = -1$, but this may also be affected by its unresolved
nature.  The M subdwarf calibration, indicating that the system is probably 
slightly more
metal-rich than -1, is more consistent with the LP 215-35 photometry 
than with the G 176-046ABC data.  
In any case, we can at least conclude that 
$[m/H] = -1.07$ measurement is consistent with the position of 
G 176-046D above the $[m/H] \sim -1.2$ sdM but below the 
$[m/H] \sim -0.6$ stars.  

The spectroscopic classification can be tested against
position in the HR diagram for VB 12\footnote{Neither of the 
other systems has a reliable trigonometric parallax measurement}.  
HD 219617 has a Hipparcos trigonometric parallax ($\pi = 12.41
\pm 2.04$ milliarcseconds).  
VB 12 has $V=16.46$, $V-I = 2.09$,
(\cite{b90}) implying $M_V = 11.93$.  These values place it
slightly above the extreme subdwarf sequence 
(\cite{m92}; \cite{g97}) as expected from the bandstrength
indices.  

The current G97 M-subdwarf metallicity scale is in good
agreement with the CLLA metallicities although the possibility remains
of systematic uncertainties at the $\pm 0.3$ dex level.  
One potential problem is the elemental abundance ratios used in the
Allard \& Hauschildt (1995) model atmospheres which form the basis of the G97
calibration. Those atmospheres are computed using scaled solar metallicities.
However, Ruan's (1991) spectroscopic analysis of VB 12 (and HD 219617) 
indicates that both stars have the expected oxygen- 
and $\alpha$-enhancement, and that changing 
the abundance ratios has a significant effect on the colors. 
Baraffe et al. (1997) have argued that the appropriate 
method of comparing stellar interior models (as well as 
stellar atmosphere models) for the M-subdwarfs is to consider
$[m/H] \approx [O/H] = [Fe/H] + [O/Fe] \approx [Fe/H] + 0.35$ for
$[Fe/H] \le -1$.  They find that the Monet 
et al. (1992) esdM are consistent with an average  $[m/H] \sim -1.3$ or $-1.5$ 
based upon the $(M_V,V-I$) HR diagram, whereas G97 found 
$[m/H] \sim -2 \pm 0.5$ for the same stars on the basis of spectroscopy.
The discrepancy between the two calibrations is thus at least 0.5 dex. 

\begin{figure}
\figurenum{3}
\plotfiddle{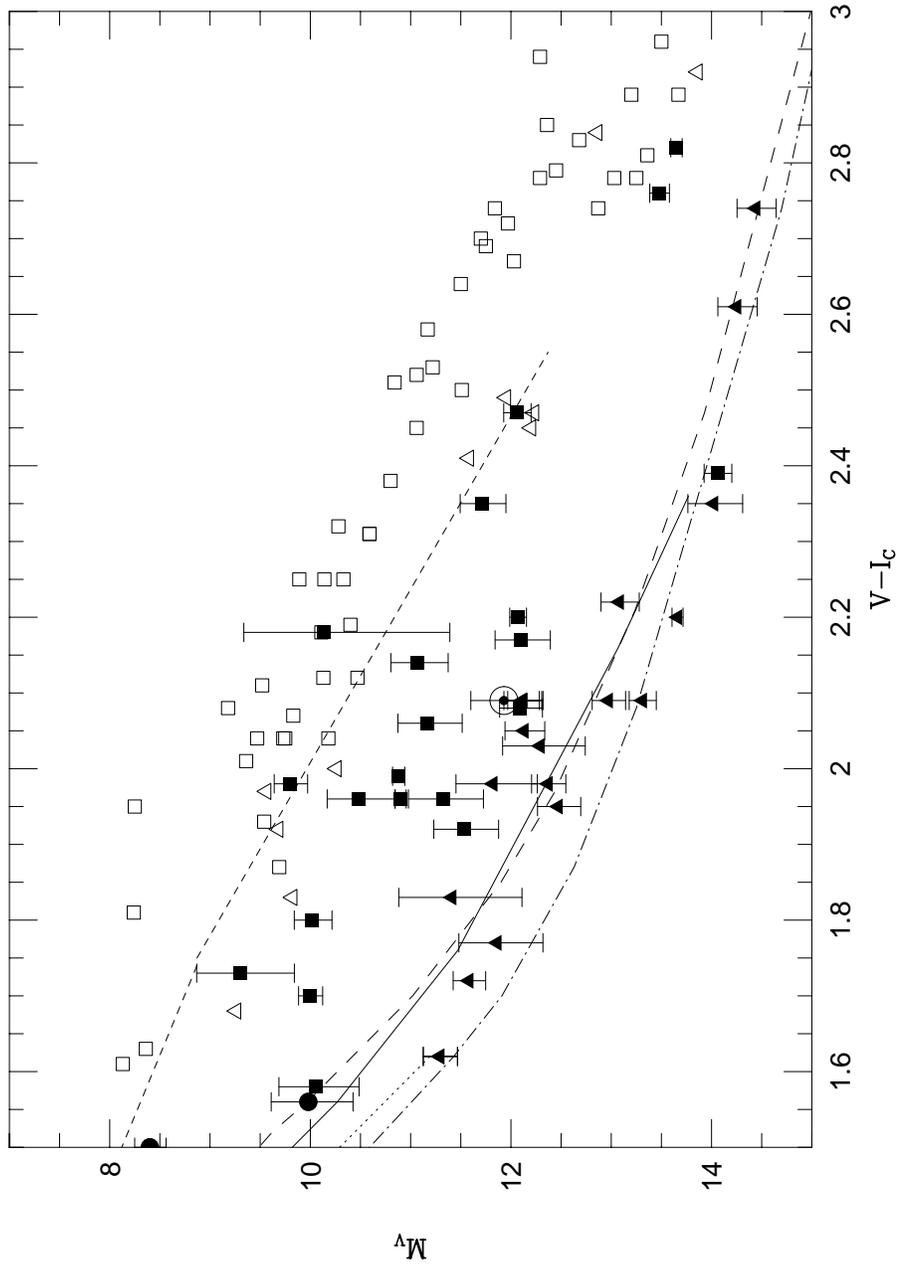}{6in}{0}{70}{70}{-230}{-30}
\caption{The HR Diagram of very low mass stars.
Shown are the esdM (solid triangles), sdM (solid squares),
high velocity M V (open triangles), and single nearby disk M V
stars (open squares).       
VB 12 is the circled point at $M_V=11.93$,$V-I = 2.09$.  
Also shown are the main sequences of three  
globular clusters:   NGC 6397 (solid line,  $[Fe/H] = -1.82$), 
M15 (dotted line, $[Fe/H] = - 2.12$), and 47 Tuc (short dashed line,
$[Fe/H] = -0.70$.   
The globular cluster main sequences use the Hipparcos-based
distances of Reid (1997a,1997b).  The recent model  
calculations of Baraffe et al. for $[m/H] = -2$ and $-1.3$
are plotted as long-dashed and dotted and as long-dashed lines
respectively.  The $-1.3$ model is in good
agreement with NGC 6397.\label{fig-newhst}}
\end{figure}

Since the publication of Gizis' analysis, Reid (1997a) and Gratton et al (1997)
have used main-sequence fitting to re-derive distances to a number of
well-known clusters, using nearby F \& G subdwarfs with high-precision Hipparcos 
parallax measurements as local calibrators. 
The resultant distances are significantly higher for the lowest abundance
clusters (such as M92, M15 and NGC 6397). 
We can compare the re-calibrated cluster  color-magnitude
diagrams against the (M$_V$, (V-I)) distribution of the local M subdwarfs.
In Figure~\ref{fig-newhst}, we 
plot the parallax subdwarfs classified by Gizis (1997). 
We readjust published globular cluster main sequences
to the distances and reddenings 
used by Reid (1997a, 1997b).  We use the clusters
NGC 6397 (\cite{cpk96}), M15 (\cite{seg96}), and 47 Tuc (\cite{seg96}).
The metallicities ($[Fe/H$) for these clusters are -1.82, -2.12,
and -0.70 respectively (\cite{cg97}).\footnote{The Baraffe et al. (1997) 
$[m/H] = -1.3$ and $-2$ tracks are also plotted.  With the new distances, the
best fitting model for NGC 6397 is now $[m/H] = -1.3$ rather than -1.5.}  
NGC 6397's main sequence passes through the esdM sequence.  
M15 lies at the bottom edge of
esdM distribution, although the HST data do not extend very far to the
red.  As in G97, 47 Tuc lies above the sdM, indicating that they have
$[m/H] < -0.7$.   Thus, an empirical calibration suggests that the G97 
M subdwarf abundance scale is consistent with the globular cluster $[Fe/H]$
scale.  

\section{Summary \label{conclusions}}

We have compared the metallicities estimated directly from spectra of
three M subdwarfs to the metallicities derived for
their FGK subdwarf companions.  We find that the metallicities
based on the Gizis (1997) spectroscopic classification system 
are consistent with the metallicities derived by Carney et al. (1994)
from high-resolution spectra.   We argue that $[m/H]$ on 
the G97 scale corresponds to $[m/H] \approx [Fe/H]$.  
    
\acknowledgments

We thank the Palomar Observatory staff for their capable support.
JEG gratefully acknowledges support by Greenstein and
Kingsley Fellowships as well as NASA grants GO-06344.01-95A and
GO-05913.01-94A.
This research has made use of the Simbad database, operated at
CDS, Strasbourg, France.



\begin{deluxetable}{lccc}
\footnotesize
\tablecaption{Spectroscopic Indices}

\tablewidth{0pt}
\tablenum{1}
\label{indices}
\tablehead{
\colhead{Band} & \colhead{S1} & \colhead{W} & \colhead{S2}
}
\startdata
TiO 5  & 7042-7046 &7126-7135 & \nl
CaH 1  & 6345-6355 &6380-6390 &6410-6420 \nl
CaH 2  & 7042-7046 &6814-6846\nl
CaH 3  & 7042-7046 &6960-6990 \nl
\enddata
\end{deluxetable}

\begin{deluxetable}{crrcrrrrrcr}
\tablewidth{0pc}
\tablenum{2}
\label{table-data}
\tablecaption{Data}
\tablehead{
\colhead{Star} & 
\colhead{V} &
\colhead{V-I$_C$} &
\colhead{Source\tablenotemark{a}} &
\colhead{Sep. ($\arcsec$)} &
\colhead{TiO5} &
\colhead{CaH1} &
\colhead{CaH2} &
\colhead{CaH3} &
\colhead{Sp. Type} &
\colhead{$[m/H]_{p}$} 
}
\startdata
G 116-009B & 18.4\phantom{0} & 2.0\phantom{0} & MR & 10.2 & 0.93 & 0.83 & 0.65 & 0.84 & sdM0.5 & -1.46 \nl
G 176-046D & 18.0\phantom{0} & 2.2\phantom{0} & MR & 4.7 & 0.76 & 0.80 & 0.63 & 0.83 & M 1.0 V & -1.07 \nl
VB 12 & 16.46 & 2.09 & B & 15\phantom{.}\phantom{0} &0.71 &  0.60 &  0.44 &  0.65 & sdM3.0 & -1.40 \nl 
\enddata
\tablenotetext{a}{Photometry sources are MR (Martin \& Rebolo 1992; 
Martin et al. 1995) and B (Bessell 1990).}   
\end{deluxetable}

\clearpage


\clearpage

\end{document}